\documentclass[aps,amssymb,amsmath,
twocolumn,showpacs,superscriptaddress,
groupedaddress,ifmbe]{revtex4-1}

\usepackage{graphicx}
\usepackage{natbib}
\usepackage{dcolumn}
\usepackage{bm}
\usepackage[table,xcdraw]{xcolor}
\usepackage{nth}
\usepackage[ruled,vlined]{algorithm2e}
\usepackage{framed}
\usepackage{listings}
\usepackage[titletoc,title]{appendix}
\usepackage{tikz}
\usepackage{lipsum}
\usepackage{hyperref}
\usepackage{multirow}
\usepackage{nicefrac}
\usepackage[table]{xcolor}
\usepackage{threeparttable}
\usepackage{comment}
\usepackage{amsmath}
\usepackage[autostyle]{csquotes}
\usepackage{float}
\usepackage{placeins}

\hypersetup{
  colorlinks,
  citecolor=red,
  linkcolor=blue,
  urlcolor=blue}

\begin{document}
\title{$N-$dimensional Coulomb$-$Sturmians with noninteger quantum numbers.}
\author{Ali Ba{\u g}c{\i}}
\email{abagci@pau.edu.tr}
\affiliation{Computational and Gravitational Physics Laboratory, Department of Physics, Faculty of Science, Pamukkale University, Denizli, Turkey}

\begin{abstract}
Coulomb$-$Sturmian functions are complete, orthonormal, and include the full spectrum of continuum states. They are restricted to integer values of quantum numbers, as imposed by boundary and orthonormality conditions. Ba{\u g}c{\i}$-$Hoggan exponential$-$type orbitals remove this restriction through a generalization to quantum number with fractional order.
\\
The differential equations for $N-$dimensional Ba{\u g}c{\i}$-$Hoggan orbitals are derived. It is demonstrated that Coulomb$-$Sturmian functions satisfy a particular case of these equations. Additionally, Guseinov's $\Psi^{\alpha}-$ETOs are identified as $N-$dimensional Coulomb$-$Sturmians with a shifted dimensional parameter $\alpha$, rather than representing an independent complete orthonormal sets of basis in a weighted Hilbert space.

\begin{description}
\item[Keywords]
Ba{\u g}c{\i}$-$Hoggan ETOs, Coulomb$-$Sturmian functions, Fractional quantum numbers
\end{description}
\end{abstract}
\maketitle

Bağcı$-$Hoggan complete and orthonormal exponential$-$type orbitals (BH$-$ETOs) \cite{1_Bagci2023} originate from the solution of the Dirac equation \cite{2_Dirac1928} in the non$-$relativistic limit. They are given as,
\begin{multline}\label{eq:1}
\mathcal{R}^{\nu}_{n^{\ast}l^{\ast}}\left(\zeta, r\right)=
\mathcal{N}^{\nu}_{n^{\ast}l^{\ast}}\left(\zeta\right)
\\
\times
\left(2\zeta r\right)^{l^{\ast}+\nu-1} e^{-\zeta r}
\mathcal{L}^{2l^{\ast}+2\nu}_{n^{\ast}-l^{\ast}-\nu}\left(2\zeta r\right)
\end{multline}
Here, $0 < \nu \leq 1$, the asterisk over the parameters, such as $x^{*}$, indicates that these parameters belong to the set of real numbers $\left( x \in \mathbb{R} \right)$. $\mathcal{N}^{\nu}_{n^{\ast}l^{\ast}}$ are normalization constants. They are given as,
\begin{align}\label{eq:2}
\mathcal{N}^{\nu}_{n^{\ast}l^{\ast}}\left(\zeta \right)=
\Bigg[
\frac
{\left(2\zeta\right)^{3}\Gamma\left[n^{\ast}-l^{\ast}-\nu+1\right]}
{\Gamma\left[n^{\ast}+l^{\ast}+\nu+1\right]}
\Bigg]^{1/2}
\end{align}
$\mathcal{L}_{q^{*}-p^{*}}^{p^{*}}$ are generalized Laguerre polynomials satisfy the following orthogonality relationship,
\begin{multline}\label{eq:3}
\int_{0}^{\infty}
\left(2\zeta r\right)^{p^{\ast}} e^{-2\zeta r}
\mathcal{L}_{q^{\ast}-p^{\ast}}^{p^{\ast}}\left(2\zeta r\right)
\mathcal{L}_{q^{\prime \ast}-p^{\prime \ast}}^{p^{\prime \ast}}\left(2\zeta r\right) dr
\\
=\frac
{\Gamma\left[q^{\ast}+1\right]}
{\Gamma\left[q^{\ast}-p^{\ast}+1\right]}
\delta_{q^{\ast}-p^{\ast},q^{\prime \ast}-p^{\prime \ast}} .
\end{multline}
BH$-$ETOs from the Kepler problem as formulated by Infeld and Hull \cite{3_Infeld1951} using the factorization method. A compact analytical representation is obtained \cite{1_Bagci2023} by extending the associated Laguerre polynomials to noninteger order using fractional calculus \cite{4_Kilbas2006}, wherein transitional Laguerre polynomials are introduced as an intermediate form between associated and generalized Laguerre polynomials. The complete analytical derivation of the BH$-$ETOs is subsequently achieved through this formalism. Analogous to Guseinov's $\Psi^{\alpha}-$ETOs \cite{5_Guseinov2002}, the BH$-$ETOs are defined in the weighted Hilbert space $L^{2}_{r^{\alpha}}(\mathbb{R}^{3})$ \cite{6_Klahn1977} as, 
\begin{multline}\label{eq:4}
\mathcal{R}^{\alpha \nu}_{n^{\ast}l^{\ast}}\left(\zeta, r\right)=
\mathcal{N}^{\alpha \nu}_{n^{\ast}l^{\ast}}\left(\zeta\right)
\\
\times
\left(2\zeta r\right)^{l^{\ast}+\nu-1} e^{-\zeta r}
\mathcal{L}^{2l^{\ast}+2\nu-\alpha}_{n^{\ast}-l^{\ast}-\nu}\left(2\zeta r\right)
\end{multline}
The function space spanned by the BH$-$ETOs transcends the standard Hilbert space, which may require the introduction of appropriate Hilbert$-$Sobolev spaces. For each class of BH$-$ETOs, a corresponding discretization scheme is specified, and the associated orbitals span a Sobolev space \cite{7_Sobolev1963}.

The differential equations for $\Psi^{\alpha}-$ETOs are first derived in agreement with Guseinov \cite{8_Guseinov2012} using the following relationships for Laguerre polynomials,
\begin{align}\label{eq:5}
\frac{d}{dx}\mathcal{L}^{p^{\ast}}_{q^{\ast}-p^{\ast}}\left(x\right)
=\mathcal{L}^{p^{\ast}+1}_{q^{\ast}-p^{\ast}-1}\left(x\right),
\end{align}
and,
\begin{multline}\label{eq:6}
x\frac{d^{2}}{dx^{2}}\mathcal{L}^{p^{\ast}}_{q^{\ast}-p^{\ast}}\left(x\right)
+\left(p^{\ast}+1-x\right)
\frac{d}{dx}\mathcal{L}^{p^{\ast}}_{q^{\ast}-p^{\ast}}\left(x\right)
\\
+\left(q^{\ast}-p^{\ast}\right)
\mathcal{L}^{p^{\ast}}_{q^{\ast}-p^{\ast}}\left(x\right)=0
\end{multline}
For standard Coulomb$-$Sturmians and $\Psi^{\alpha}-$ETOs, $p^{\ast}=p$, $q^{\ast}=q$, $\left\lbrace p,q \right\rbrace \in \mathbb{N}_{0}$. The resulting form however, corresponds to an improper separation of variable$-$dependent and constant terms, which obscures the true properties of Coulomb$-$Sturmians. The source of this issue lies in the premature application of one$-$center expansion methods (Guseinov, 2002 \cite{5_Guseinov2002}), prior to deriving the differential equation (Guseinov, 2012 \cite{8_Guseinov2012}). $\Psi^{\alpha}-$ETOs have been subjected to considerable criticism \cite{9_Weniger2007, 10_Weniger2008, 11_Weniger2012, 12_Bagci2025, 13_Bagci2025}, some of these concerns have mainly addressed the Hilbert space representation and convergence characteristics \cite{9_Weniger2007, 10_Weniger2008, 11_Weniger2012}, predicated on the notion that the parameter $\alpha$ is an observable or variational parameter. It has recently been established by the author \cite{12_Bagci2025, 13_Bagci2025} that $\alpha$ is neither observable nor suitable as a variational parameter.
\\
The present letter reports that Guseinov's approach to construct a set of complete orthonormal functions in a weighted Hilbert space constitutes just another route for representing Coulomb$-$Sturmians in $N-$dimensional space \cite{14_Avery2002, 15_Aquilanti2001, 16_Coletti2013}. The sole distinction here, is that the three$-$dimensional case is used as reference. It is demonstrated that BH$-$ETOs do not exhibit such deficiencies and provide the proper generalization of Coulomb$-$Sturmians to noninteger quantum numbers. This is achieved by considering the non$-$relativistic limit of the Dirac$-$Coulomb equation, where the solutions are the BH$-$ETOs. 

$\Psi^{\alpha}-$ETOs emerge as a special case of the Eq. (\ref{eq:4}) when $\nu=1$. Upon substituting $p=2l+2-\alpha$, $q=n+l+1-\alpha$, and $x=2\zeta r$ into the differential equation given in Eq. (\ref{eq:6}) for Laguerre polynomial, straightforward algebra confirms that the differential equation for $\Psi^{\alpha}$-ETOs reduces to the form reported in \cite{8_Guseinov2012}:\\
$\mathcal{R}^{\alpha}_{nl}\left(x\right) \equiv \mathcal{R}^{\alpha}_{nl}$,
\begin{multline}\label{eq:7}
x\frac{d^2}{dx^2}\mathcal{R}^{\alpha}_{nl}
+\left(3-\alpha\right)\frac{d}{dx}\mathcal{R}^{\alpha}_{nl}
+\left[
n+\left(1-\alpha\right)
\left(\frac{1}{2}-\frac{l}{x}\right)
\right.
\\
\left.
-\frac{l\left(l+1\right)}{x}
-\frac{x}{4}
\right]\mathcal{R}^{\alpha}_{nl}=0 .
\end{multline}
The differential equation given in Eq. (\ref{eq:7}) cannot be straightforwardly generalized to the relativistic case, as it relies on an improper separation of variable$-$dependent and constant terms. Any consistent rearrangement must be compatible with the Coulomb$-$Sturmian differential equation and with the spin$-$orbit coupling between the orbital angular momentum $\hat{\mathbf{L}}$ and the spin operator $\hat{\sigma}$, which together determine the total angular momentum $\hat{\mathbf{J}} = \hat{\mathbf{L}} + \tfrac{1}{2}\hat{\sigma}$, as follows from the symmetry and the integrability of the Dirac Hamiltonian for an electron in a Coulomb potential. The integrable structure of the Dirac$-$Coulomb problem is characterized by commuting operators $\left\{ \hat{H}_{D}, \hat{J}^{2}, \hat{J}_{z}, \hat{K} \right\}$, where $\hat{K}$ is the Dirac invariant \cite{17_Bagci2025} (see also references therein). Specifically, the Dirac invariant operator is defined as $\hat{K} = \beta(\sigma \cdot \hat{L} + \hbar\mathbb{I})$, which in atomic units ($\hbar = 1$) reduces to $\hat{K} = \beta(\sigma \cdot \hat{L} + \mathbb{I})$, where $\beta$ is the standard Dirac $\beta$ matrix and $\mathbb{I}$ is the identity operator in spinor space. The eigenvalues of $\hat{K}$, denoted $\kappa$, take the discrete values $\kappa = \pm(j + \frac{1}{2})$, where the orbital angular momentum quantum number is given by $l = -\kappa - 1$ for $\kappa < 0$ (corresponding to $j = l + \frac{1}{2}$) and $l = \kappa$ for $\kappa > 0$ (corresponding to $j = l - \frac{1}{2}$), representing the spin$-$orbit coupled states with total angular momentum quantum number $j$.\\
The correct formulation must therefore in nonrelativistic limit preserve the spin$-$orbit coupling structure of the Dirac$-$Coulomb problem which is satisfied by the Coulomb$-$Sturmians. The Eq. (\ref{eq:7}) may be rewritten below as,
\begin{multline}\label{eq:8}
x\frac{d^2}{dx^2}\mathcal{R}^{\alpha}_{nl}
+\left(3-\alpha\right)\frac{d}{dx}\mathcal{R}^{\alpha}_{nl}
+\left[
n+\left(1-\alpha\right)
\left(\frac{1}{x}+\frac{1}{2}\right)
\right.
\\
\left.
-\frac{l\left(l+1-\alpha\right)}{x}
-\frac{x}{4}
\right]\mathcal{R}^{\alpha}_{nl}=0 .
\end{multline}
By changing the variable as $r=\frac{x}{2\zeta}$ and and applying subsequent algebraic simplification, the differential equations reduce to the following form:
\begin{multline}\label{eq:9}
-\frac{1}{r}\frac{d}{dr}\mathcal{R}^{\alpha}_{nl}
-\frac{1}{\left(3-\alpha\right)}\frac{d^{2}}{dr^{2}}\mathcal{R}^{\alpha}_{nl}
+\frac
{l\left(l+2-\alpha\right)}
{\left(3-\alpha\right)r^{2}}
\\
-\frac
{\zeta\left(2n+1-\alpha\right)}
{\left(3-\alpha\right)r}\mathcal{R}^{\alpha}_{nl}
-E\mathcal{R}^{\alpha}_{nl}=0,
\end{multline}
where, $\left(3-\alpha\right)E=-\zeta^{2}$. Collecting the radial derivative terms, the differential equation reads:
\begin{multline}\label{eq:10}
\Bigg[
-\frac{1}{\left(3-\alpha\right)r^{3-\alpha}}\frac{\partial}{\partial r}
\Big(
r^{3-\alpha}\frac{\partial}{\partial r}
\Big)
+\frac
{l\left(l+2-\alpha\right)}
{\left(3-\alpha\right)r^{2}}
\\
-\frac
{\zeta\left(2n+1-\alpha\right)}
{\left(3-\alpha\right)r}
-E\Bigg]\mathcal{R}^{\alpha}_{nl}=0.
\end{multline}
The Eq. (\ref{eq:10}) is identical to that given in \cite{14_Avery2002,15_Aquilanti2001}, where $\alpha=4-N$, $N$ denotes the dimensionality. Analogously, the differential equations for $N-$dimensional $BH-$ETOs are obtained as,
\begin{multline}\label{eq:11}
\Bigg[
-\frac{1}{\left(3-\alpha\right)r^{3-\alpha}}\frac{\partial}{\partial r}
\Big(
r^{3-\alpha}\frac{\partial}{\partial r}
\Big)
\\
+\frac
{\left(l^{\ast}+\nu-1\right)\left(l^{\ast}+\nu+1-\alpha\right)}
{\left(3-\alpha\right)r^{2}}
\\
-\frac
{\zeta\left(2n+1-\alpha\right)}
{\left(3-\alpha\right)r}
-E\Bigg]\mathcal{R}^{\nu \alpha}_{nl}=0.
\end{multline}
The angular momentum eigenvalues in the generalized $N-$dimensional case can be expressed as $\lambda_{l^{\ast},\nu}^{(N)} = (l^{\ast} + \nu - 1)(l^{\ast} + \nu + N - 3)$. To establish the connection with standard hyperspherical harmonics, the effective angular momentum quantum number is introduced as, $l^{\prime \ast} = l^{\ast} + \nu - 1$. The eigenvalues then assume the canonical form $\lambda_{l^{\prime \ast}}^{(N)} = l^{\prime \ast}(l^{\prime \ast} + N - 2)$, similarly to the eigenvalue of the Laplace$-$Beltrami operator acting on hyperspherical harmonics $Y_{l^{\prime \ast}}^{(N)}(\Omega_N)$ defined on the $(N-1)-$dimensional hypersphere \cite{14_Avery2002}. In the limiting case $\nu = 1$, one also recovers $\lambda_{l^{\ast}}^{(N)} = l^{\ast}(l^{\ast} + N - 2)$, thereby confirming that the BH$-$ETOs with $\nu = 1$ coincide with conventional $N$-dimensional Coulomb$-$Sturmians for $l^{\ast}=l$, $l \in \mathbb{N}_{0}$. Consider the three$-$dimensional case ($N = 3$, $\alpha = 1$), the eigenvalues reduce to $\lambda_{l^{\ast},\nu}^{(3)} = (l^{\ast} + \nu - 1)(l^{\ast} + \nu)$. Setting $l^{\prime \ast} = l^{\ast} + \nu - 1$ again, yields the familiar form to $l^{\prime \ast}(l^{\ast} + 1)$, standard spherical harmonics with fractional angular momentum quantum numbers. This fractional generalization necessitates the extension of Gegenbauer polynomials $C_n^{(\lambda)}(x)$ to noninteger indices or may be a revision in the kinetic energy operator as indicated in \cite{17_Bagci2025}.
\section*{Acknowledgement}
The authors declares no conflict of interest.
\pagebreak


\begin{thebibliography}{plainnat}

\bibitem{1_Bagci2023} Ba{\u g}c{\i} A, Hoggan PE (2023) \textit{Complete and orthonormal sets of exponential$-$type orbitals with non$-$integer quantum numbers.} J. Phys. A: Math. Theor. \textbf{56}(33): 335205. doi: \url{https://doi.org/10.1088/1751-8121/ace6e2}

\bibitem{2_Dirac1928} Dirac PAM (1928) \textit{The quantum theory of the electron.} Proc. R. Soc. Lond. Ser. A$-$Contain. Pap. Math. Phys. Character \textbf{117}(778): 610--624. doi: \url{https://doi.org/10.1098/rspa.1928.0023}

\bibitem{3_Infeld1951} Infeld L, Hull TE (1951) \textit{The Factorization Method.} Rev. Mod. Phys. \textbf{23}(1): 21--68. doi: \url{https://link.aps.org/doi/10.1103/RevModPhys.23.21}

\bibitem{4_Kilbas2006} Kilbas AA, Srivastava HM and Trujillo JJ (2006) \textit{Theory and Applications of Fractional Differential Equations} (North$-$Holland Mathematics Studies, Elsevier)

\bibitem{5_Guseinov2002} Guseinov II (2002) \textit{New complete orthonormal sets of exponential-type orbitals and their application to translation of Slater orbitals.} Int. J. Quant. Chem. \textbf{90}(1): 114--118. doi: \url{https://doi.org/10.1002/qua.927}

\bibitem{6_Klahn1977} Klahn B, Bingel WA (1977) \textit{The convergence of the Rayleigh$-$Ritz Method in quantum chemistry.} Theor. Chim. Acta \textbf{44}(1): 27--43. doi: \url{https://doi.org/10.1007/BF00548027}

\bibitem{7_Sobolev1963} Sobolev SL (1963) \textit{Applications of Functional Analysis in Mathematical Physics.} (American Physical Society, Rhode Island).

\bibitem{8_Guseinov2012} Guseinov II (2012) \textit{New Complete Orthonormal Sets of Exponential$-$Type Orbitals in Standard Convention and Their Origin.} Bull.Chem.Soc.Jpn \textbf{85}(12): 1386--1389. doi: \url{https://doi.org/10.1246/bcsj.20120207} 

\bibitem{9_Weniger2007} Weniger EJ (2007) \textit{Extended Comment on ``One$-$Range Addition Theorems for Coulomb Interaction Potential and Its Derivatives" by I. I. Guseinov (Chem. Phys. Vol. 309 (2005), pp. 209 - 213).}  arXiv:0704.1088v3 [math-ph]. doi: \url{ 	
https://doi.org/10.48550/arXiv.0704.1088}

\bibitem{10_Weniger2008} Weniger EJ (2008) \textit{On the analyticity of Laguerre series.} J. Phys. A: Math. Theor. \textbf{41}(42): 425207. doi: \url{https://doi.org/10.1088/1751-8113/41/42/425207}

\bibitem{11_Weniger2012} Weniger EJ (2012) \textit{On the mathematical nature of Guseinov’s rearranged one-range addition theorems for Slater-type functions.} J. Math. Chem. \textbf{50}(1): 17--81. doi: \url{https://doi.org/10.1007/s10910-011-9914-4}

\bibitem{12_Bagci2025} Ba{\u g}c{\i} A, Hoggan PE (2025) \textit{New atomic orbital functions. Complete and orthonormal sets of ETOs with non$-$integer quantum numbers. Results for He$-$like atoms.} Adv. Quant. Chem. \textbf{92}: 51--69. doi: \url{https://doi.org/10.1016/bs.aiq.2025.07.004}

\bibitem{13_Bagci2025} Ba{\u g}c{\i} A, Hoggan PE (2025) \textit{Complete and orthonormal sets of exponential$-$type orbitals with non$-$integer quantum numbers. On the results for many$-$electron atoms using Roothaan’s LCAO method.} Adv. Quant. Chem. \textbf{92}: 71--92. doi: \url{https://doi.org/10.1016/bs.aiq.2025.07.006}

\bibitem{14_Avery2002} Avery J (2002) \textit{Hyperspherical Harmonics and Generalized Sturmians.} In Progress in Theoretical Chemistry and Physics Series. (Springer Dordrecht). doi: \url{https://doi.org/10.1007/0-306-46944-8}

\bibitem{15_Aquilanti2001} Aquilanti V, Cavalli S, Coletti C, Domenico D snd Grossi G (2001) \textit{Hyperspherical harmonics as Sturmian orbitals in momentum space: A systematic approach to the few$-$body Coulomb problem.} Int. Rev. Phys. Chem. \textbf{20}(4): 673--709. doi: \url{https://doi.org/10.1080/01442350110075926}

\bibitem{16_Coletti2013} Coletti C, Calderini D and Aquilanti V (2013) \textit{d$-$Dimensional Kepler$-$Coulomb Sturmians and Hyperspherical Harmonics as Complete Orthonormal Atomic and Molecular Orbitals.} Adv. Quant. Chem. \textbf{67}: 73--127. doi: \url{https://doi.org/10.1016/B978-0-12-411544-6.00005-4}

\bibitem{17_Bagci2025} Ba{\u g}c{\i} A, Hoggan PE (2025) \textit{Relativistic exponential$-$type spinor orbitals and their use in many$-$electron Dirac equation solution.} Adv. Quant. Chem. \textbf{91}: 67--93. doi: \url{https://doi.org/10.1016/bs.aiq.2025.03.007}

\end{thebibliography}
\end{document}